\documentclass{amsart}
\usepackage{amscd,amssymb,subfigure,hyperref, epsfig}
\usepackage[arrow,matrix,graph,frame,poly,arc,tips]{xy}
\usepackage{amsmath}

\theoremstyle{plain}
\newtheorem{thm}[subsection]{Theorem}
\newtheorem{defn}[subsection]{Definition}

\newtheorem{prop}[subsection]{Proposition}
\newtheorem{cor}[subsection]{Corollary}

\theoremstyle{definition}

\numberwithin{equation}{section}

\newcommand{\F}{{\mathbb F}}

\newcommand{\Z}{{\mathbb Z}}

\newcommand{\R}{{\mathbb R}}

\newcommand{\Fqstar}{\F_q^{\,*}}
\newcommand{\conv}{\mbox{\rm conv}}
\renewcommand{\P}{{\mathbb P}}

\begin{document}
\title[On $m$-dimensional toric codes]{On $m$-dimensional toric codes}
\author{John Little}
\address{Department of Mathematics and Computer Science,
College of the Holy Cross,  Worcester, MA 01610}
\email{\href{mailto:little@mathcs.holycross.edu}{little@mathcs.holycross.edu}}
\urladdr{\href{http://mathcs.holycross.edu/~little/homepage.html/}%
{http://mathcs.holycross.edu/\~{}little/homepage.html}}
\author{Ryan Schwarz}
\address{Department of Mathematics and Computer Science,
College of the Holy Cross,  Worcester, MA 01610}

\subjclass[2000]{Primary 94B27; Secondary 52B20, 14M25}

\keywords{coding theory, toric code, Vandermonde matrix}

\date{July 25, 2005}

\begin{abstract}
\noindent
Toric codes are a class of $m$-dimensional cyclic codes introduced
recently by J. Hansen in \cite{jh}, \cite{jh1}, and studied
in \cite{j}, \cite{dgv}, \cite{ls}.
They may be defined as evaluation codes obtained from monomials
corresponding to integer lattice points in an integral 
convex polytope $P \subseteq \R^m$.  As such, they are
in a sense a natural extension of Reed-Solomon codes.  Several articles
cited above use intersection theory on toric surfaces to derive bounds
on the minimum distance of some toric codes with $m = 2$.
In this paper, we will provide a more elementary approach that
applies equally well to many toric codes for all $m \ge 2$.
Our methods are based on a sort of multivariate generalization of
Vandermonde determinants that has also been used in the study of 
multivariate polynomial interpolation. We use these Vandermonde
determinants to determine the minimum distance of toric codes
from rectangular polytopes and simplices.  We also prove a general 
result showing that if there is a unimodular integer affine 
transformation taking one polytope $P_1$ to a second polytope $P_2$, 
then the corresponding toric codes are monomially equivalent (hence have
the same parameters).  We use this to begin a classification
of two-dimensional cyclic toric codes with small dimension.
\end{abstract}
\maketitle

\section{Introduction}

In \cite{jh}, J. Hansen introduced the notion of a toric
code. Let $P \subset \R^m$ be an integral convex
polytope (the convex hull of some set of integer lattice points).
We suggest \cite{z} as a good general reference for the geometry of polytopes.
Suppose that $P \cap \Z^m$ 
is properly contained in the rectangular box $[0,q-2]^m$
(which we denote $\Box_{q-1})$, for some prime power $q$.  
Then a toric code is obtained
by evaluating linear combinations of the monomials with exponent vector in $P \cap \Z^m$ 
at some subset (usually all) of the points of $(\Fqstar)^m$. 
We formalize this in the following definition.

\begin{defn}
Let $\F_q$ be a finite field with primitive element
$\alpha$. For $f \in \Z^m$ with $0\le f_i \le q-2$
for all $i$, let $p_f = (\alpha^{f_1},\ldots,\alpha^{f_m})$
in $(\Fqstar)^m$. For each $e = (e_1,\ldots,e_m) \in P \cap \Z^m$, 
let $x^e$ be the corresponding monomial and write 
$$(p_f)^e = (\alpha^{f_1})^{e_1}\cdots (\alpha^{f_m})^{e_m}.$$
The toric code $C_P(\F_q)$ over the field
$\F_q$ associated to $P$ is the linear code of 
block length $n = (q-1)^m$ with generator matrix 
$$G = ((p_f)^e),$$
where the rows are indexed by the $e \in P \cap \Z^m$, and the columns are
indexed by the $p_f \in (\Fqstar)^m$.
In other words, letting $L = \mbox{\rm Span}\{x^e : e \in P\cap \Z^m\}$, we
define the evaluation mapping
\begin{eqnarray*} \mbox{\rm ev} : L & \to & \F_q^{\,(q-1)^m}\\
                                  g & \mapsto & (g(p_f) : f \in (\Fqstar)^m)
\end{eqnarray*}
Then $C_P = \mbox{\rm ev}(L)$.
If the field is clear from the context, we will often omit it
in the notation and simply write $C_P$.  The matrix $G$
will be called the standard generator matrix for the toric code.
\end{defn}

Because of the close connection between integral polytopes
and the theory of toric varieties, Hansen and others
have proposed techniques from algebraic geometry such as intersection theory
on algebraic surfaces and higher dimensional varieties to study
toric codes and their parameters.  The articles
\cite{jh}, \cite{jh1}, \cite{j}, and \cite{ls} have used
this approach.  

In this article, for the most part, we use a more elementary
viewpoint, based on the observation that the square submatrices
of the standard generator matrix of a toric code are examples of
a multivariate generalization of the familiar univariate 
Vandermonde matrices.  These multivariate Vandermonde
matrices have been studied by a number of 
different techniques in the context of multivariate polynomial
interpolation.  The literature there is truly vast because
of the many applications of interpolation in numerical
analysis and other parts of applied mathematics. We
direct the reader to the bibliography of 
\cite{gs}.  To our knowledge, these techniques have not 
been used before in coding theory in this form.  

Our contributions are as follows.  In \S 2, we begin
with a lower bound on the minimum distance of toric codes
based on Vandermonde determinants (Proposition 2.1).  We use this
to study the minimum distances of toric codes from 
rectangular polytopes and simplices (see Theorems 2.4, 2.5, 2.9). 
We do this by identifying special configurations of points in 
$(\Fqstar)^m$ for which the Vandermonde determinant
is nonzero.  In the context of polynomial interpolation,
such sets are called {\it poised sets} for the interpolation
problem using linear combinations of monomials corresponding
to the lattice points in some polytope.  (They are the sets
for which the interpolation problem has a unique solution for
all function values assigned at those points.)  Our methods
here are suggested by the interpolation-theoretic computations
in \cite{cl}.

All of the 2-dimensional
examples from \cite{jh} and \cite{jh1} can be handled with
our methods.  But in fact our theorems are {\it more general}
since they apply for all $m\ge 2$, not only the case $m = 2$.

In \S 3 we prove the general statement that lattice equivalent
polytopes $P_1,P_2$ yield monomially equivalent toric codes
$C_{P_1}, C_{P_2}$.  We apply this result to consider a 
classification of $m = 2$ toric codes up to monomial equivalence
for small $k$.  

Finally, some comments about the utility of toric codes are
probably in order.  In the case $m = 1$, a toric code is just a 
Reed-Solomon code since $P$ is a line segment in 
$[0,q-2] \subset \R$ with integer
endpoints.  Higher dimensional toric codes are 
in a sense a natural extension of Reed-Solomon codes and have
many similar properties.  For instance, it is easy to 
see that they are all $m$-dimensional cyclic codes (see \cite{dgv}). So 
one might hope that toric codes exist having similarly good parameters.
And indeed, \cite{j} contains a number of examples showing that 
toric codes can have very good parameters, equaling or bettering
the best known minimum distance for a given $n,k$ in \cite{b}.
Not all toric codes perform this well, however.  Our main results
on toric codes from rectangular polytopes and simplices show,
in fact, that their minimum distances are often quite small for their
dimensions.  It is an interesting problem, we believe, to 
determine criteria concerning which polytopes yield good toric codes.

We will use this notation in the following sections.
Suppose that $P \subset \Box_{q-1} \subset \R^m$
is an integral convex polytope.  We will write $\#(P)$
for the number of integer lattice points in $P$ (that 
is, $\#(P) = |P \cap \Z^m|$).  We will write 
$$P \cap \Z^m = \{e(i) : i = 1, \ldots, \#(P)\}$$
for the set of those integer lattice points.  
For any set $S \subset \R^n$, $\mbox{conv}(S)$ denotes
the convex hull of $S$.

\section{Minimum distances via Vandermonde matrices}

 We begin
by describing the Vandermonde matrices involved here.  
Using the notation introduced
in \S 1, let $P$ be an integral convex polytope, and
suppose $P \cap \Z^m = \{e(i) : i = 1, \ldots, \#(P)\}$,
listed in some particular order.
Let $S = \{p_j : j = 1, \ldots, \#(P)\}$ 
be any set of $\#(P)$ points in $(\Fqstar)^m$, also ordered. Then
$V(P; S)$, the 
{\it Vandermonde matrix} associated to $P$ and $S$, is the 
$\#(P)\times \#(P)$ matrix 
$$V(P; S) = \left(p_j^{e(i)}\right),$$
where we use the standard multi-index notation $p_j^{e(i)}$
to indicate the value of the monomial $x^{e(i)}$ at the point
$p_j$.  For example, if $P = \mbox{conv}\{(0,0),(2,0),(0,2)\}$
in $\R^2$, and $S = \{(x_j,y_j)\}$ is any set of 6 points in 
$(\Fqstar)^2$, for one particular choice of ordering of the 
lattice points in $P$, we have
$$V(P; S) = \begin{pmatrix} 1 & 1 & 1 & 1 & 1 & 1\\
                  x_1 & x_2 & x_3 & x_4 & x_5 & x_6 \\
                  y_1 & y_2 & y_3 & y_4 & y_5 & y_6 \\
                  x_1^2 & x_2^2 & x_3^2 & x_4^2 & x_5^2 & x_6^2 \\
                  x_1y_1& x_2y_2 & x_3y_3 & x_4y_4 & x_5y_5 & x_6y_6 \\
                  y_1^2 & y_2^2 & y_3^2 & y_4^2 & y_5^2 & y_6^2 \\
                  \end{pmatrix}\leqno(1)$$

Since we assume $P \subset \Box_{q-1}$, the 
monomials $x^{e(i)}$ define linearly independent functions 
on $(\Fqstar)^m$ and we can also view the $V(P; S)$ as square
submatrices of the standard generator matrix for the toric code
$C_P = C_P(\F_q)$.  Our first observation is that Vandermonde determinants
may be used to bound the minimum distance of a toric code.

\begin{prop} Let $P \subset \R^m$ be an integral convex polytope.  Let $d$
be a positive integer and assume that in every set 
$T \subset (\Fqstar)^m$ with $|T| = (q-1)^m - d + 1$
there exists some $S \subset T$ with $|S| = \#(P)$
such that $\det V(P; S) \ne 0$.
Then the minimum distance satisfies $d(C_P) \ge d$.
\end{prop}

\begin{proof}  All codewords of $C_P$ are linear combinations of the 
rows of the $\#(P) \times (q-1)^m$ standard generator matrix.
If a codeword has zeroes in the locations corresponding to the subset 
$T \subset (\Fqstar)^m$, then by looking at the entries corresponding
to $S \subset T$ we get a system of $\#(P)$ homogeneous linear equations
in the coefficients of the linear combination, whose 
matrix is $V(P; S)$.  By hypothesis, this matrix is nonsingular,
so all the coefficients in the linear combination must be zero.  
Since this is true for all $T$, all nonzero codewords of $C_P$ 
have at most $(q-1)^m - d$ zero entries, which implies $d(C_P) \ge d$.
\end{proof}

We will apply this proposition by identifying particular configurations
of points $S$ for which $\det V(P ; S) \ne 0$, based on the particular 
monomials appearing in $P$.  From examples such as
(1) above, it should be relatively clear that giving characterizations
of $S$ such that $\det V(P; S) = 0$ or 
$\det V(P; S) \ne 0$ is difficult in general.  Indeed,
even in the context of multivariate polynomial interpolation, 
only partial results in special cases are well understood.

We will begin with the case where $P$ is a rectangular polytope
in $\R^m$
(special cases are rectangles in the plane and rectangular solids
in $\R^3$).  For these polytopes, the extension from the case 
$m = 2$ to general $m \ge 2$ is almost immediate, so for notational
simplicity, we will treat only the case $m = 2$ in detail.  

Let $P_{k,\ell}$ be the rectangle 
$P_{k,\ell} = \mbox{conv}\{(0,0), (k,0), (0,\ell), (k,\ell)\}$.
Note that $\#(P_{k,\ell}) = (k+1)(\ell + 1)$.  We will call 
any set $S$ of $(k+1)(\ell + 1)$ points in $(\Fqstar)^2$
consisting of $(\ell + 1)$ distinct points on each of $(k+1)$
distinct vertical lines $x = a_i$ a $(k+1) \times (\ell + 1)$
{\it configuration}.  

When we construct $V(P_{k,\ell}; S)$ for a $(k+1) \times (\ell + 1)$
configuration we get a matrix as in the following proposition.

\begin{prop} 
Suppose $A=(c_{ij})$ is an $a \times a$ matrix and 
$B_1 ,B_2, \ldots, B_a$ are $b \times b$ matrices.  Let 
{\bf M} be the $ab \times ab$ block matrix:
$${\bf M}=\begin{pmatrix} c_{11}B_1& c_{12}B_2 & \cdots & c_{1a}B_a\\
                   \vdots   & \vdots    &        & \vdots   \\
                   c_{a1}B_1& c_{a2}B_2 & \cdots & c_{aa}B_a
           \end{pmatrix}$$
Then $\det({\bf M})=\pm \det(A)^b \det(B_1) \det (B_2) \cdots 
\det(B_a)$.
\end{prop}

The matrix ${\bf M}$ is similar to a tensor product matrix, but the $B_i$
may be different matrices, so this construction is somewhat more general.

\begin{proof}
If $\det(A)=0$ or $\det(B_i) = 0$ for some $i$, then $\det({\bf M})=0$ as well.  So, 
we assume that $\det(A)\ne 0$ and $\det(B_i) \ne 0$ for all $i$.
In order to find the determinant of ${\bf M}$, we may
transform ${\bf M}$ into a block upper triangular matrix 
using blockwise row operations, obtaining
a matrix ${\bf M'}$ in the following form:
$${\bf M'}=\begin{pmatrix} c_{11}'B_1&         &      &       &    *     \\
                   0        &c_{22}'B_2&      &       &         \\
                   \vdots   & 0       &\ddots&       &         \\
                   \vdots   & \vdots  &\ddots&\ddots &         \\
                   0        &  0      &\cdots& 0     &c_{aa}'B_a 
            \end{pmatrix},$$
in which the $c_{ii}'$'s are the same as the entries obtained
by the corresponding row operations applied to the matrix $A$.  
Now we have a block-upper 
triangular matrix.  This implies that the determinant 
of the matrix is the product of the determinants of the diagonal block 
entries.   Thus,
\begin{eqnarray*}
\ \det({\bf M'}) & = & (c_{11}')^b\det(B_1) \cdot (c_{22}')^b\det(B_2)\cdots (c_{aa}')^b\det(B_a)  \\
\ & = & (c_{11}' \cdots c_{aa}')^b\det(B_1) \det(B_2)\cdots \det(B_a)\\
\ & = & \det(A')^b \det(B_1) \det (B_2) \cdots \det(B_a)
\end{eqnarray*}
We know that $\det(A')=\pm \det(A)$ (some row interchanges might have
been necessary in the reduction to upper triangular form).  So, 
we can substitute to yield: 
$$ \det({\bf M})  = \pm \det(A)^b \det(B_1) \det (B_2) \cdots \det(B_a),$$
which is what we wanted to show.
\end{proof}

When $S$ is a $(k+1) \times (\ell + 1)$ configuration, consisting
of $\ell + 1$ distinct points on each of $k + 1$ distinct
lines $x = a_u$, $u = 1, \ldots, k + 1$,   
then $V(P_{k,\ell}; S)$ has the form given in the proposition,
where $A$ is the ordinary $(k+1) \times (k+1)$
univariate Vandermonde matrix of the 
$x = a_u$, and $B_u$ is the $(\ell + 1)\times (\ell + 1)$
univariate Vandermonde matrix of
the $y$-coordinates of the points on $x = a_u$.
Hence we obtain the following consequence.

\begin{cor}
Let $S$ be a $(k+1) \times (\ell + 1)$ configuration
in $(\Fqstar)^2$.  Then $$\det V(P_{k,\ell}; S) \ne 0.$$
\end{cor}

\begin{proof}  This follows from the factorization of
the univariate Vandermonde determinants as products of differences
of the $x$- and $y$-values at the points in $S$.
\end{proof}

We are now ready to prove the first major result of this
section.

\begin{thm} Let $k,\ell < q - 1$ so that $P_{k,\ell} \subset \Box_{q-1} \subset \R^2$. 
Then the minimum distance of the two-dimensional toric code
$C_{P_{k,\ell}}$ is 
$$d(C_{P_{k,\ell}}) = (q-1)^2 - (k+\ell)(q-1) + k\ell = ((q-1) - k)((q-1) - \ell).$$
\end{thm}

\begin{proof}
We write $d = d(C_{P_{k,\ell}})$.  
In order to show equality, we will show that both
$d \le (q-1)^2-(k+\ell) (q-1) + k\ell$ and 
$d\ge (q-1)^2-(k+\ell)(q-1) + k\ell$.  We start with the former.
In $L=\mbox{Span}\{1, x, x^2, \dots, x^k, y, xy, x^2y, \dots , x^ky, \ldots, x^ky^\ell\}$
consider a polynomial $p(x,y)=q_1(x)q_2(y)$ where $q_1(x)$ and $q_2(y)$ factor 
completely: 
$$p(x,y)=(x-a_1)(x-a_2)\cdots(x-a_k)(y-b_1)(y-b_2)\cdots (y-b_\ell),$$
for some distinct $a_u$ and $b_v$ in $\Fqstar$
This means that the codeword $\mbox{ev}(p)$
has zeros in the positions corresponding to the points along $k$ distinct vertical lines and 
$\ell$ distinct horizontal lines in $(\Fqstar)^2$.  We see that $\mbox{ev}(p)$ has 
$(k+\ell)(q-1) - k\ell$ zeroes.  Accordingly, we have $d \le (q-1)^2-(k+\ell)(q-1)+k\ell$.

Now we show $d\ge (q-1)^2-(k+\ell)(q-1)+k\ell$.  
By Corollary 2.3 and Proposition 2.1, it suffices to show
that every set $T$ of size $(k+\ell)(q-1)-k\ell+1$ in $(\Fqstar)^2$ contains
a $(k+1)\times (\ell+1)$ configuration $S$.  
Since $q - 1 > k$, we see that $(k+\ell)(q-1)-k\ell + 1 > \ell (q-1)$.  
So, by the pigeonhole principle, some vertical 
line $x=a_1$ contains $\ell+1$ distinct points from $T$.  There are at most 
$(q-1)-(\ell+1)=q-\ell-2$ other points on that line.  Therefore, there are at least 
$(k+\ell-1)(q-1)-k\ell + 1$ other points in $T$ not on $x=a_1$.  To repeat this
argument to find the rest of the configuration $S$, we must show that for all $1\le j \le k$,
$(k + \ell -j)(q-1) - k\ell + 1 > \ell (q - 1 - j)$.  To do this, we subtract $\ell (q - 1 - j)$ 
and perform some arithmetic:  
$$
(k + \ell - j)(q-1) - k\ell + 1 - \ell (q - 1 - j) = (k - j)((q-1) - \ell) + 1 > 0.$$
So, after we have found the first $j$ 
sets of $\ell + 1$ points on vertical lines in $T$, 
there are still enough additional points left in $T$ so 
that we can find the next set of $\ell+1$ points. After $k+1$ steps, we
have a complete
$(k+1) \times (\ell + 1)$ configuration $S \subset T$.   
\end{proof}

The method used in the proof of Theorem 2.4 extends without difficulty to 
toric codes constructed from $k_1\times k_2 \times \cdots \times k_m$
rectangular polytopes $P_{k_1,\ldots,k_m} \subset \R^m$ for all $m \ge 2$.  
The result is as follows.

\begin{thm} Let $k_1,\ldots,k_m$ be small enough so that 
$P_{k_1,\ldots,k_m} \subset \Box_{q-1} \subset \R^m$. 
Then the minimum distance of the $m$-dimensional toric code
$C_{P_{k_1,\ldots,k_m}}$ is 
$$d(C_{P_{k_1,\ldots,k_m}}) = \prod_{i=1}^m ((q-1) - k_i).$$
\end{thm}

We next turn to the toric codes $C_{P_\ell(m)}$ for $P_\ell(m)$ an $m$-dimensional
simplex of the form
$$P_\ell(m) = \mbox{conv}\{{\bf 0},\ell {\bf e}_1,\ldots, \ell {\bf e}_m\},$$
where the ${\bf e}_i$ are the standard basis vectors in $\R^m$.  The monomials
corresponding to the ${m + \ell\choose\ell}$ integer lattice points in 
$P_\ell(m)$ are all of the monomials
in $m$ variables of total degree $\le \ell$.  Naturally enough, the corresponding 
Vandermonde matrices arise in the study of multivariate polynomial interpolation
using polynomials of bounded total degree.  The next recursive definition gives the 
special configurations of points where we will be able to compute
Vandermonde determinants for the $P_\ell(m)$.

\begin{defn} 
If $m = 1$, an $\ell$th order {\bf simplicial configuration} 
is any collection of ${1 + \ell \choose \ell}$
distinct points in $\Fqstar$.
For $m \ge 2$, 
we will say that a collection $S$ of ${m + \ell\choose\ell}$ points 
in $(\Fqstar)^m$ is an $m$-dimensional 
$\ell$th order {\bf simplicial configuration} if the following 
conditions hold:
\begin{enumerate}
\item For some $i$, $1\le i \le m$, there are hyperplanes 
$x_i = a_1, x_i = a_2, \ldots, x_i = a_{\ell+1}$
such that for each $1 \le j \le \ell+1$, $S$ contains exactly
${m - 1 + j - 1  \choose j - 1}$ points with $x_i = a_j$.  (Note that 
$${m + \ell \choose \ell} = \sum_{j = 1}^{\ell + 1} {m - 1 + j - 1  \choose j - 1} \leqno(2)$$
by a standard binomial coefficient identity.)
\item For each $j$, $1 \le j \le \ell + 1$,  
the points in $x_i = a_j$ form an $(m-1)$-dimensional simplicial
configuration of order $j-1$.  
\end{enumerate}
\end{defn}

We call these special configurations of points simplicial configurations
because they mimic, to an extent, the arrangement of the integer
lattice points in the corresponding simplex $P_\ell(m)$.  
For instance, Figure 1 shows a $2$-dimensional simplicial
configuration of order 2 in $(\F_8^{\,*})^2$.  It consists 
of six points. (We write $\alpha$ for a primitive
element in the field $\F_8$.)
 Note that there is $1 = {1 + 0 \choose 0}$ point in $S$ on the
line $x_1 = a_1 = \alpha^4$, $2 = {1 + 1\choose 1}$ points on 
the line $x_1 = a_2 = \alpha^3$, and 
$3 = {1 + 2 \choose 2}$ on the line $x_1 = a_3 = \alpha$.

\begin{center}
\begin{picture}(200,205)(0,0)
\put(20,20){\vector(1,0){165}}
\put(20,20){\vector(0,1){165}}
\put(43,43){\circle*{5}}
\put(43,89){\circle*{5}}
\put(43,158){\circle*{5}}
\put(89,89){\circle*{5}}
\put(89,66){\circle*{5}}
\put(112,43){\circle*{5}}

\put(0,0){$1$}
\put(43,0){$\alpha$}
\put(66,0){$\alpha^2$}
\put(89,0){$\alpha ^3$}
\put(112,0){$\alpha ^4$}
\put(135,0){$\alpha ^5$}
\put(158,0){$\alpha^6$}
\put(45,17){\line(0,1){5}}
\put(68,17){\line(0,1){5}}
\put(91,17){\line(0,1){5}}
\put(114,17){\line(0,1){5}}
\put(137,17){\line(0,1){5}}
\put(160,17){\line(0,1){5}}

\put(0,43){$\alpha$}
\put(0,66){$\alpha^2$}
\put(0,89){$\alpha^3$}
\put(0,112){$\alpha^4$}
\put(0,135){$\alpha^5$}
\put(0,158){$\alpha^6$}
\put(17,43){\line(1,0){5}}
\put(17,66){\line(1,0){5}}
\put(17,89){\line(1,0){5}}
\put(17,112){\line(1,0){5}}
\put(17,135){\line(1,0){5}}
\put(17,158){\line(1,0){5}}
\end{picture}\\
$\quad$\\
Figure 1.  A 2-dimensional simplicial configuration $S$ of order 2.
\end{center}

In order to state our next result, a sort of recurrence relation
for the Vandermonde determinants $\det V(P_\ell(m); S)$ where $S$ is a 
an $m$-dimensional simplicial configuration, we introduce
some notation.  

Let $S$ be an $m$-dimensional $\ell$th order simplicial 
configuration consisting of ${m + \ell \choose \ell}$
points, in hyperplanes $x_m = a_1, \ldots, x_m = a_{\ell+1}$.
Write $S = S'\cup S''$ where $S'$ is the union
of the points in $x_i = a_1, \ldots, a_\ell$, and $S''$
is the set of points in $x_i = a_{\ell + 1}$.  Also,
let $\pi : \F_q^{\,m} \to \F_q^{\,m-1}$ be the projection on the
first $m - 1$ coordinates.  By the definition, it follows
that both $S'$ and $\pi(S'')$ are themselves simplicial
configurations, with $S'$ of dimension $m$ and order $\ell - 1$,
and $\pi(S'')$ of dimension $m - 1$ and order $\ell$.

\begin{thm} Let $P_\ell(m)$ be as above and let $S$ be
an $\ell$th order simplicial configuration of ${m + \ell \choose \ell}$
points as in the paragraph above.  
Then writing $p = (p_1,\ldots,p_m)$ for points $p\in (\Fqstar)^m$, 
$$\det V(P_\ell(m); S) = \pm \prod_{p \in S'}(p_m - a_{\ell +1}) 
\det V(P_{\ell-1}(m); S') \det V (P_{\ell}(m-1); \pi(S'')).$$
\end{thm}

Before we give the proof of this theorem, we will do 
two things.  First, we give an example to illustrate
what the theorem is saying. The idea for this computation
comes from \cite{cl}, where corresponding sets of points in 
$\R^m$ are identified as poised sets for interpolation
by polynomials of degree bounded bounded by $\ell$. 
Consider all polynomials of degree $\le 2$ in three 
variables and the Vandermonde matrix $V(P_2(3); S)$.  
For notational simplicity,
write points in a 3-dimensional simplicial 
configuration $S \subset (\Fqstar)^3$ of order 2
as $(x_i,y_i,z_i)$, for $i = 1,\ldots, 10 = {3 + 2\choose 2}$.  
Here $S'$ consists of the first four points in $S$, and $S''$
consists of the other six points.
Under the hypothesis that 
$S$ is a simplicial configuration, we have
$z_5 = z_6 = \cdots  = z_{10} = c$ for some $c = a_3$.  
Noting this, but ignoring other equalities between the 
coordinates, we see $V(P_2(3); S) = $
$$
\begin{pmatrix} 1 & 1 & 1 & 1 & 1 & 1 & 1 & 1 & 1 & 1\\
               x_1& x_2& x_3& x_4& x_5& x_6& x_7& x_8& x_9& x_{10}\\
               y_1& y_2& y_3& y_4& y_5& y_6& y_7& y_8& y_9& y_{10}\\
               z_1& z_2& z_3& z_4& c& c& c& c& c& c\\
               x_1^2& x_2^2& x_3^2& x_4^2& x_5^2& x_6^2& x_7^2& x_8^2& x_9^2& x_{10}^2\\
               x_1y_1& x_2y_2& x_3y_3& x_4y_4& x_5y_5& x_6y_6& x_7y_7& x_8y_8& x_9y_9& x_{10}y_{10}\\
               y_1^2& y_2^2& y_3^2& y_4^2& y_5^2& y_6^2& y_7^2& y_8^2& y_9^2& y_{10}^2\\
               x_1z_1& x_2z_2& x_3z_3& x_4z_4& x_5c& x_6c& x_7c& x_8c& x_9c& x_{10}c\\
               y_1z_1& y_2z_2& y_3z_3& y_4z_4& y_5c& y_6c& y_7c& y_8c& y_9c& y_{10}c\\
               z_1^2& z_2^2& z_3^2& z_4^2& c^2& c^2& c^2& c^2& c^2& c^2
               \end{pmatrix}.$$
To evaluate the determinant of this matrix, we perform row operations 
to introduce zeroes.  First subtract $c$ times row 4 from row 10, then $c$ times row 3 from row 9,
$c$ times row 2 from row 8, and finally $c$ times row 1 from row 4.  
After rearranging rows, we find a matrix with a block of zeroes:
$$\begin{pmatrix}
               z_1-c& \cdots & z_4-c& 0& 0& 0& 0& 0& 0\\
               x_1(z_1-c)& \cdots & x_4(z_4-c)& 0& 0& 0& 0& 0& 0\\
               y_1(z_1-c)& \cdots & y_4(z_4-c)& 0& 0& 0& 0& 0& 0\\
               z_1(z_1-c)& \cdots & z_4(z_4-c)& 0& 0& 0& 0& 0& 0\\
               1 & \cdots & 1 & 1 & 1 & 1 & 1 & 1 & 1\\
               x_1& \cdots & x_4& x_5& x_6& x_7& x_8& x_9& x_{10}\\
               y_1& \cdots & y_4& y_5& y_6& y_7& y_8& y_9& y_{10}\\
               x_1^2& \cdots & x_4^2& x_5^2& x_6^2& x_7^2& x_8^2& x_9^2& x_{10}^2\\
               x_1y_1& \cdots & x_4y_4& x_5y_5& x_6y_6& x_7y_7& x_8y_8& x_9y_9& x_{10}y_{10}\\
               y_1^2& \cdots & y_4^2& y_5^2& y_6^2& y_7^2& y_8^2& y_9^2& y_{10}^2
\end{pmatrix}.\leqno(3)$$
Up to a sign, the determinant of $V(P_2(3); S)$ is therefore equal to 
the product of 
$$\det \begin{pmatrix} z_1 - c & z_2 - c & z_3 - c & z_4 - c \\
x_1(z_1 - c) & x_2(z_2 - c) & x_2(z_3 - c) & x_4(z_4 - c) \\
y_1(z_1 - c) & y_2(z_2 - c) & y_3(z_3 - c) & y_4(z_4 - c) \\
z_1(z_1 - c) & z_2(z_2 - c) & z_3(z_3 - c) & z_4(z_4 - c)
\end{pmatrix},$$
which equals 
$$(z_1 - c)(z_2 - c)(z_3 - c)(z_4 - c) \det V(P_1(3),S'),$$
and the determinant of the lower right $6\times 6$ block in (3),
which is $V(P_2(2),\pi(S''))$.  Hence
$$\det V(P_2(3); S) = \pm \prod_{j=1}^4(z_j - c) \det V(P_1(3),S')\det V(P_2(2),\pi(S'')),$$
as in the statement of the theorem.

Second, we note the following
immediate consequence of the theorem.

\begin{cor} Let $P_\ell(m)$ be as above and let $S$ be
an $\ell$th order simplicial configuration of ${m + \ell \choose \ell}$
points.  Then $\det V(P_\ell(m); S) \ne 0$.
\end{cor}

\begin{proof}
This is seen easily by a double induction on $m$ and $\ell$.  In the base
cases $m = 1$, $\ell$ arbitrary, we have an ordinary univariate Vandermonde
determinant, which is nonzero by the definition of a 
simplicial configuration.  For the induction step, the recurrence
given in the theorem then establishes the corollary since all the 
factors are nonzero under the hypothesis that $S$ is a
simplicial configuration.
\end{proof}

We will now give a proof of Theorem 2.7, following the methods
from \cite{cl} used in the example above.

\begin{proof}  
Since $S$ is simplicial, the ${m + \ell - 1\choose \ell - 1}$ points
in $S'' \subset S$ all have the same $x_m$-coordinate, say $x_m = a_{\ell + 1} = c$.  
For each pair of monomials $x^e$ and $x^e x_m$ corresponding to
points in $P_\ell(m)$, we perform a row operation on the Vandermonde
matrix, subtracting $c$ times the row of $V(P_\ell(m),S)$ 
for $x^e$ from the row for $x^e x_m$ in {\it decreasing} order 
by the degree in $x_m$.  After rearranging rows to put all the 
zeroes created by these operations in the upper right block,
the lower right block in the columns corresponding to $S''$
is the matrix $V(P_\ell(m-1),\pi(S''))$.  In the upper left block 
(in the columns corresponding to $S'$), all the entries in a column
are divisible by one of the $(p_m - c)$ for $p\in S'$.  Factoring
out those factors from each column, the matrix that is left
is $V(P_{\ell-1}(m),S')$, and the statement of the theorem follows.
\end{proof}

We note that if we apply the recurrence relation repeatedly,
another corollary of Theorem 2.7 is a closed
formula for $\det V(P_\ell(m); S)$ for $S$ a simplicial 
configuration in terms of univariate Vandermonde determinants.
We will not need this formula, so we leave the derivation 
of its exact form as an exercise.  

We will now use Corollary 2.8 to establish the minimum distances
of the toric codes $C_{P_\ell(m)}$.

\begin{thm}  Let $\ell < q - 1$, and let $P_\ell(m)$ be
the simplex in $\R^m$ defined above.  Then the minimum
distance of the toric code $C_{P_\ell(m)}$ is given
by 
$$d(C_{P_\ell(m)}) = (q-1)^m - \ell(q-1)^{m-1}.$$
\end{thm}

\begin{proof}
Let $d = d(C_{P_\ell(m)})$.
As in the proof of Theorem 2.4 above, we show both inequalities
$d \le (q-1)^m - \ell(q-1)^{m-1}$ and $d \ge (q-1)^m - \ell(q-1)^{m-1}$
hold.  The first of these follows as in Theorem 2.4, since the completely 
reducible polynomials
$$p(x_m) = (x_m - a_1)\cdots (x_m - a_{\ell})$$
for $a_i$ distinct in $\Fqstar$ are contained in 
$L = \mbox{Span}\{x^{e} : e\in P_\ell(m)\}$.  Such a polynomial
has zeroes at all the $\F_q$-rational points with nonzero coordinates
on the union of the hyperplanes $x_m = a_i$.  There are $\ell(q-1)^{m-1}$
such points, so $d \le (q-1)^m - \ell(q-1)^{m-1}$ as claimed.

To establish the reverse inequality, by Corollary 2.8 and 
Proposition 2.1, it suffices to show that every 
set $T$ of $\ell (q-1)^{m-1} + 1$ points
contains an $m$-dimensional simplicial configuration $S$ of order $\ell$.  
By the pigeonhole principle, $T$ contains some set of 
$\ell(q-1)^{m-2} + 1$ points with the same $x_m$-coordinate
$x_m = a_{\ell + 1}$.
By an easy induction on $m$, it follows that 
$$\ell(q-1)^{m-2} + 1 > {m - 1 + \ell \choose \ell}$$
(the first term on the right of (2)) for all $m \ge 2$ (and $q - 1 > \ell$, of course)
There are at least $(\ell-1)(q-1)^{m-1} + 1$ points
in $T$ that {\it do not} lie on $x_m = a_{\ell + 1}$.  Hence
we can apply the same pigeonhole principle argument repeatedly
(or argue by induction on $\ell$) to see that 
$T$ contains 
$$\sum_{j=1}^{\ell + 1} {m - 1 + j - 1\choose j - 1} = {m + \ell \choose \ell}$$
points making up an $m$-dimensional $\ell$th order simplicial configuration.
\end{proof}

We also have the following consequence for more general simplices.
Let $\ell_i \ge 1$ for all $i$ and define
$$P_{\ell_1,\ldots,\ell_m} = \mbox{conv}\{{\bf 0}, \ell_1 {\bf e}_1, \ldots, \ell_m {\bf e}_m\},$$
where again the ${\bf e}_i$ are the standard basis vectors in $\R^m$.

\begin{cor} If $P_{\ell_1,\ldots,\ell_m} \subset \Box_{q-1} \subset \R^m$, and 
$\ell = \max_i \ell_i$, then 
$$d(C_{P_{\ell_1,\ldots,\ell_m}}) = (q-1)^m - \ell(q-1)^{m-1}.$$
\end{cor}

\begin{proof}  By definition, $P_{\ell_1,\ldots,\ell_m} \subseteq P_\ell(m)$.
Hence $C_{P_{\ell_1,\ldots,\ell_m}}$ is a subcode of $C_{P_\ell(m)}$ and 
$d(C_{P_{\ell_1,\ldots,\ell_m}}) \ge d(C_{P_\ell(m)}) = (q-1)^m - \ell(q-1)^{m-1}$.
But if $\ell = \ell_{i_0}$, then
$C_{P_{\ell_1,\ldots,\ell_m}}$ also contains codewords of weight exactly
$(q-1)^m - \ell(q-1)^{m-1}$ obtained from evaluation of completely reducible 
polynomials 
$$p(x_i) = (x_{i_0} - a_1)\cdots (x_{i_0} - a_{\ell})$$
for some $i$ and distinct $a_j \in \Fqstar$.  This establishes the corollary.
\end{proof}

The same sort of reasoning applies to any toric code from a polytope
$P \subset P_\ell(m)$ that contains one complete edge 
$\conv\{{\bf 0},\ell{\bf e}_i\}$ or 
$\conv\{\ell{\bf e}_i,\ell{\bf e}_j\}$ of the simplex
$P \subset P_\ell(m)$ and gives the same minimum distance for $C_P$.

\section{Classification of toric codes}

In this section we will begin by stating and proving a theorem
guaranteeing that two toric codes have the same parameters.  
We begin by introducing some terminology.

\begin{defn}
Let $C_1$ and $C_2$ be two codes of block length $n$ and dimension $k$ over
$\F_q$.  Let $G_1$ be a generator matrix for $C_1$.  Then 
$C_1$ and $C_2$ are said to be
{\bf monomially equivalent} if there is an invertible $n\times n$ diagonal
matrix $\Delta$ and an $n\times n$ permutation matrix $\Pi$ such that
$$G_2 = G_1 \Delta \Pi$$
is a generator matrix for $C_2$.
\end{defn}

It is easy to see that monomial equivalence is actually an 
equivalence relation on codes since a product $\Pi\Delta$
equals $\Delta'\Pi$ for another invertible diagonal matrix 
$\Delta'$.  It is also 
a direct consequence of the definition that monomially
equivalent codes $C_1$ and $C_2$ have the same dimension 
and the same minimum distance
(indeed, the same full weight enumerator). 

Next we turn to a natural notion of equivalence for polytopes.
Recall that an {\it affine transformation} of $\R^m$ is a mapping
of the form $T({\bf x}) = M{\bf x} + \lambda$, where $\lambda$
is a fixed vector and $M$ is an $m\times m$ matrix.
The affine mappings $T$ where $M, \lambda$ have integer
entries and $M \in \mbox{GL}(m,\Z)$ (so $\det(M) = \pm 1$) are precisely the 
bijective affine mappings from the integer lattice $\Z^m$ to itself. 

\begin{defn}
We will say that two integral convex polytopes $P_1$ and $P_2$ in $\R^m$ are 
are {\bf lattice equivalent} if there exists an 
invertible integer affine transformation $T$ as above
such that $T(P_1) = P_2$.
\end{defn}

This brings us to our next theorem which relates the two concepts we have
just defined.

\begin{thm}
If two polytopes $P_1$ and $P_2$ are lattice equivalent, then the
toric codes $C_{P_1}$ and $C_{P_2}$ are monomially equivalent.
\end{thm} 

\begin{proof}
Suppose we have two lattice equivalent polytopes $P_1$ and $P_2$.
Both $P_1$ and $P_2$ contain integer lattice points corresponding to 
monomials of the form $x^e$ where $e \in \Z^m$.  By our hypothesis on 
$P_1$ and $P_2$, there exists an invertible integer transformation 
$$T({\bf x}) = M({\bf x}) + \lambda$$
such that $T(P_1)=P_2$ and $M$ is an element of 
$GL(m,\Z)$ so $\det(M) = \pm 1$.  
Hence $\#(P_1) = \#(P_2)$.  Let
$P_1\cap \Z^m = \{e(i): i = 1, \ldots,\#(P_1)\}$.
So, $C_{P_1}$ is spanned by $\mbox{ev}(x^{e(i)})$ for $1 \le i \le n$, and
similarly $C_{P_2}$ is spanned by $\mbox{ev}(x^{T(e(i))})$.
Write $\alpha$ for a primitive element in $\Fqstar$.
Let $e(i) \in P_1\cap \Z^m$ and define
$\alpha^f = (\alpha^{f_1}, \ldots, \alpha^{f_m}) \in (\Fqstar)^m$.  
The component of 
$\mbox{ev}(x^{e(i)}) \in C_{P_1}$ corresponding to $\alpha^f$ is 
$\alpha^{\langle e(i), f \rangle}$, where $\langle e(i), f\rangle$
is the usual dot product.  The corresponding entry in the 
codeword $\mbox{ev}(x^{T(e(i))})$ in $C_{P_2}$ is
is $\alpha^{\langle T(e(i)), f\rangle}$.
This can be rewritten as
$$\alpha^{\langle M e(i) + \lambda, f\rangle }
  = \alpha^{\langle M e(i), f \rangle } \cdot
    \alpha^{\langle \lambda, f\rangle}$$
The second term of the product is not dependent on $e(i)$.  These nonzero
scalars are the diagonal entries in the matrix $\Delta$ as in the definition
of monomially equivalent codes.  By a standard property of dot products,
$$\alpha^{\langle M e(i), f\rangle} =
 \alpha^{\langle e(i), M^t f\rangle}.$$
The transposed matrix
$M^t$ also defines a bijective mapping from $\Z^m$ to $\Z^m$ since 
$\det(M^t)=\det(M)=\pm 1$.  Now we must show that $M^t$ induces a 
permutation of $(\Fqstar)^m$.  Suppose $M^t f \equiv
M^t g \pmod {q-1}$.  Since $\det(M^t) = \pm 1 \neq 0$, we know
that $M^t$ is invertible and $(M^t)^{-1}$ is also an integer matrix.
So, we can multiply by $(M^t)^{-1}$ on the left.  
Hence, $f \equiv g\pmod {q-1}$ and
$M^t$ defines a permutation of the points
$\alpha^f$, as desired.  Note that
$M^t$ permutes all of the codewords in the same way.  This 
gives the permutation matrix $\Pi$.  Hence
$C_{P_1}$ is monomially equivalent to $C_{P_2}$.
\end{proof}

In the remainder of this section, we will show how this result
leads to a complete classification for toric codes with 
$m = 2$ and $k \le 5$.  The classification could also
be continued, of course, using a census of lattice equivalence 
classes of lattice polytopes with given $\#(P)$. 
 
\begin{prop}             
Every toric surface code $C_P$ with $k=2$ 
is monomially equivalent to the toric code
$C_{P_2}$ for $P_2 = \conv\{(0,0),(1,0)\}$.
\end{prop}

\begin{proof}
Let $e(1), e(2) \in \Z^2$ be the integer lattice points in $P$.
We can use a translation to map $e(1)$ to $(0,0)$.
Then let $e(2)=(a,b) \in \Z^2$.  
By convexity we have that gcd$(a,b)=1$, since
otherwise there would be additional integer lattice 
points on the line from $e(1)$ to $e(2)$ and $k = \#(P)$ would be greater than 2.  
Since $\gcd(a,b) = 1$, there exist integers $r,s$ such that $ra + sb = 1$, and 
this implies that there exists an invertible integer matrix 
$M = \begin{pmatrix} r & s\\ -b & a\end{pmatrix}$ 
such that
$M \begin{pmatrix} a\\ b \end{pmatrix}=\begin{pmatrix} 1\\ 0\end{pmatrix}.$ 
Hence there is an affine equivalence between $P$ and $P_2$.  By Theorem 3.1, this 
completes the proof.
\end{proof}

Next, we wish to find a ``nice'' lattice polygon
in each possible lattice equivalence class with $\#(P) = 3,4,5$.  
One way is to add additional points to $P_2$.  Using Pick's Theorem:
$A(P) = \#(P) + \frac{1}{2}\partial(P) - 1,$
(where $\partial(P)$ is the number of lattice points in the boundary of $P$)
then eliminating lattice equivalent polygons, we obtain the following.

\begin{thm} Every toric surface code with $3 \le k \le 5$ is monomially equivalent 
to one constructed from one of the following polygons.  
\begin{center}
\begin{picture}(340,100)(0,0)
\put(20,20){\vector(1,0){130}}
\put(20,20){\vector(0,1){70}}
\put(180,20){\vector(1,0){130}}
\put(180,20){\vector(0,1){70}}

\put(20,20){\circle*{5}}
\put(70,20){\circle*{5}}
\put(120,20){\circle*{5}}
\put(180,20){\circle*{5}}
\put(180,70){\circle*{5}}
\put(230,20){\circle*{5}}

\put(20,0){$1$}
\put(70,0){$x$}
\put(120,0){$x^2$}
\put(180,0){$1$}
\put(230,0){$x$}
\put(280,0){$x^2$}
\put(282,17){\line(0,1){5}}

\put(0,20){$1$}
\put(0,70){$y$}
\put(160,20){$1$}
\put(160,70){$y$}
\put(17,70){\line(1,0){5}}
\put(45,80){$P_3^{(1)}$}
\put(204,80){$P_3^{(2)}$}

\thicklines
\put(20,20){\line(1,0){100}}
\put(180,20){\line(1,0){50}}
\put(180,20){\line(0,1){50}}
\put(180,70){\line(1,-1){51}}
\end{picture}\\
\vglue 5pt
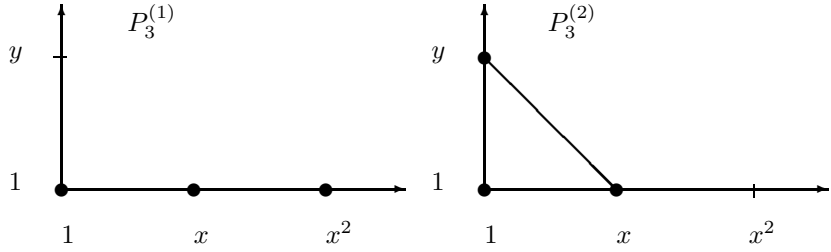
{\rm Figure 2.  Polygons yielding toric codes with $k = 3$.}
\end{center}

\begin{center}
\begin{picture}(400,290)(0,0)
\put(225,53){\vector(1,0){110}}
\put(225,53){\vector(0,1){100}}
\put(225,53){\vector(-1,0){40}}
\put(225,53){\vector(0,-1){40}}
\put(216,42){$1$}
\put(254,42){$x$}
\put(287,42){$x^2$}
\put(320,42){$x^3$}
\put(190,58){$x^{-1}$}
\put(213,86){$y$}
\put(213,119){$y^2$}
\put(228,20){$y^{-1}$}
\put(192,20){\circle*{5}}
\put(225,86){\circle*{5}}
\put(225,53){\circle*{5}}
\put(258,53){\circle*{5}}
\thicklines
\put(192,20){\line(1,2){33}}
\put(192,20){\line(2,1){65}}
\put(225,86){\line(1,-1){33}}
\put(235,155){$P_4^{(4)}$}
\thinlines

\put(60,53){\vector(1,0){100}}
\put(60,53){\vector(0,1){100}}
\put(51,42){$1$}
\put(90,42){$x$}
\put(123,42){$x^2$}
\put(48,86){$y$}
\put(48,119){$y^2$}
\put(60,53){\circle*{5}}
\put(60,86){\circle*{5}}
\put(93,53){\circle*{5}}
\put(93,86){\circle*{5}}
\thicklines
\put(60,53){\line(0,1){33}}
\put(60,53){\line(1,0){33}}
\put(93,86){\line(-1,0){33}}
\put(93,86){\line(0,-1){33}}
\thinlines
\put(70,155){$P^{(3)}_4$}

\put(60,190){\vector(1,0){100}}
\put(60,190){\vector(0,1){100}}
\put(51,179){$1$}
\put(90,179){$x$}
\put(123,179){$x^2$}
\put(156,179){$x^3$}
\put(48,223){$y$}
\put(48,256){$y^2$}
\put(60,190){\circle*{5}}
\put(93,190){\circle*{5}}
\put(126,190){\circle*{5}}
\put(159,190){\circle*{5}}
\thicklines
\put(60,190){\line(1,0){100}}
\thinlines
\put(235,290){$P_4^{(2)}$}
\put(225,190){\vector(1,0){100}}
\put(225,190){\vector(0,1){100}}
\put(216,179){$1$}
\put(254,179){$x$}
\put(287,179){$x^2$}
\put(213,223){$y$}
\put(213,256){$y^2$}
\put(225,223){\circle*{5}}
\put(225,190){\circle*{5}}
\put(258,190){\circle*{5}}
\put(291,190){\circle*{5}}
\thicklines
\put(225,190){\line(0,1){33}}
\put(225,190){\line(1,0){66}}
\put(225,223){\line(2,-1){66}}
\thinlines
\put(70,290){$P_4^{(1)}$}
\end{picture}\\
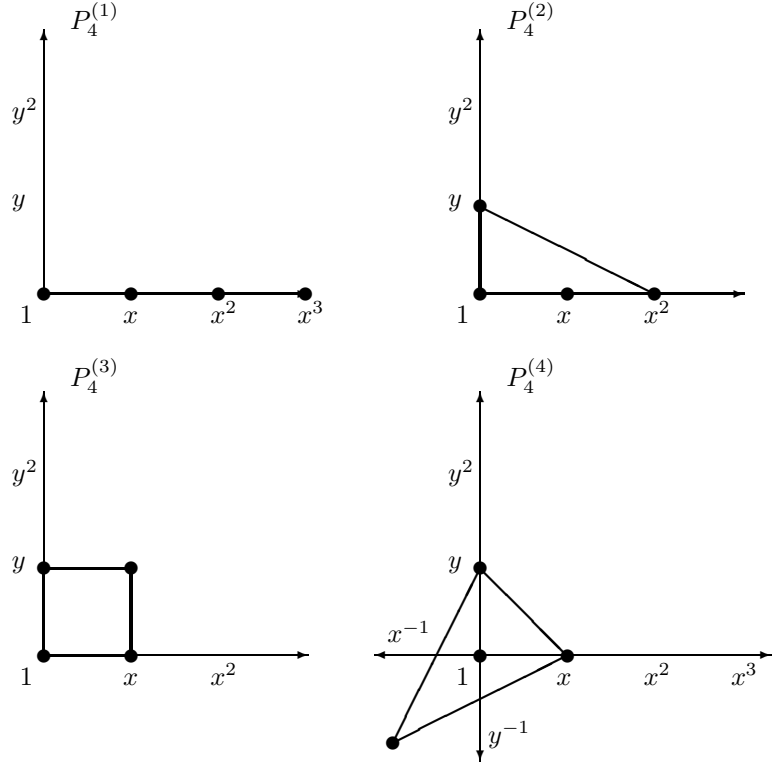
{\rm Figure 3.  Polygons yielding toric codes with $k = 4$.}
\end{center}

\begin{center}
\begin{picture}(400,200)(0,0)
\put(150,53){\vector(1,0){80}}
\put(150,53){\vector(0,1){80}}
\put(150,53){\vector(-1,0){40}}
\put(150,53){\vector(0,-1){40}}
\put(139,44){$1$}
\put(167,41){$x$}
\put(187,41){$x^2$}
\put(207,41){$x^3$}
\put(139,72){$y$}
\put(139,92){$y^2$}
\put(139,112){$y^3$}
\put(131,31){$y^{-1}$}
\put(150,53){\circle*{5}}
\put(150,73){\circle*{5}}
\put(150,33){\circle*{5}}
\put(170,53){\circle*{5}}
\put(190,53){\circle*{5}}
\thicklines
\put(150,33){\line(0,1){40}}
\put(150,33){\line(2,1){40}}
\put(150,73){\line(2,-1){40}}
\put(170,120){$P_5^{(5)}$}
\thinlines

\put(280,53){\vector(1,0){80}}
\put(280,53){\vector(0,1){80}}
\put(280,53){\vector(-1,0){40}}
\put(280,53){\vector(0,-1){40}}
\put(269,41){$1$}
\put(297,41){$x$}
\put(317,41){$x^2$}
\put(337,41){$x^3$}
\put(257,55){$x^{-1}$}
\put(269,72){$y$}
\put(269,92){$y^2$}
\put(269,112){$y^3$}
\put(285,29){$y^{-1}$}
\put(280,53){\circle*{5}}
\put(280,73){\circle*{5}}
\put(280,93){\circle*{5}}
\put(300,53){\circle*{5}}
\put(260,33){\circle*{5}}
\thicklines
\put(260,33){\line(2,1){40}}
\put(260,33){\line(1,3){20}}
\put(280,93){\line(1,-2){20}}
\put(300,120){$P_5^{(6)}$}
\thinlines

\put(20,53){\vector(1,0){80}}
\put(20,53){\vector(0,1){80}}
\put(9,41){$1$}
\put(37,41){$x$}
\put(57,41){$x^2$}
\put(77,41){$x^3$}
\put(9,72){$y$}
\put(9,92){$y^2$}
\put(9,112){$y^3$}
\put(20,53){\circle*{5}}
\put(20,73){\circle*{5}}
\put(40,53){\circle*{5}}
\put(40,73){\circle*{5}}
\put(60,93){\circle*{5}}
\thicklines
\put(20,53){\line(0,1){20}}
\put(20,53){\line(1,0){20}}
\put(40,53){\line(1,2){20}}
\put(20,73){\line(2,1){40}}
\put(40,120){$P_5^{(4)}$}
\thinlines

\put(20,150){\vector(1,0){80}}
\put(20,150){\vector(0,1){60}}
\put(9,138){$1$}
\put(37,138){$x$}
\put(57,138){$x^2$}
\put(77,138){$x^3$}
\put(97,138){$x^4$}
\put(9,169){$y$}
\put(9,189){$y^2$}
\put(20,150){\circle*{5}}
\put(40,150){\circle*{5}}
\put(60,150){\circle*{5}}
\put(80,150){\circle*{5}}
\put(100,150){\circle*{5}}
\thicklines
\put(20,150){\line(1,0){80}}
\put(40,200){$P_5^{(1)}$}
\thinlines

\put(150,150){\vector(1,0){80}}
\put(150,150){\vector(0,1){60}}
\put(139,138){$1$}
\put(167,138){$x$}
\put(187,138){$x^2$}
\put(207,138){$x^3$}
\put(139,169){$y$}
\put(139,189){$y^2$}
\put(150,150){\circle*{5}}
\put(170,150){\circle*{5}}
\put(190,150){\circle*{5}}
\put(150,170){\circle*{5}}
\put(210,150){\circle*{5}}
\thicklines
\put(150,150){\line(1,0){60}}
\put(150,150){\line(0,1){20}}
\put(150,170){\line(3,-1){60}}
\put(170,200){$P_5^{(2)}$}
\thinlines

\put(280,150){\vector(1,0){80}}
\put(280,150){\vector(0,1){60}}
\put(269,138){$1$}
\put(297,138){$x$}
\put(317,138){$x^2$}
\put(337,138){$x^3$}
\put(269,169){$y$}
\put(269,189){$y^2$}
\put(280,150){\circle*{5}}
\put(280,170){\circle*{5}}
\put(300,150){\circle*{5}}
\put(300,170){\circle*{5}}
\put(320,150){\circle*{5}}
\thicklines
\put(280,150){\line(0,1){20}}
\put(280,150){\line(1,0){40}}
\put(280,170){\line(1,0){20}}
\put(300,170){\line(1,-1){20}}
\put(300,200){$P_5^{(3)}$}
\thinlines
\end{picture}\\
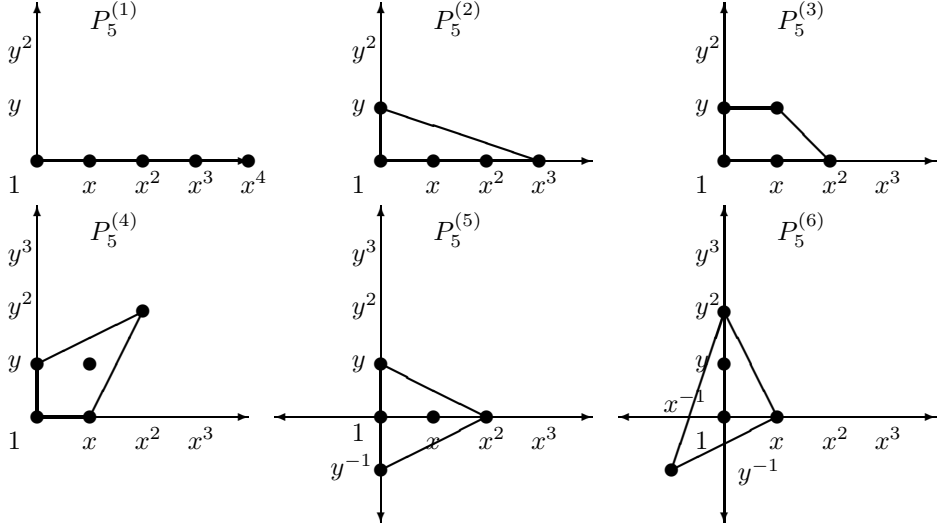
{\rm Figure 4.  Polygons yielding toric codes with $k = 5$.}
\end{center}
\end{thm}

The final step in our classification is to show that no two of the 
toric surface codes constructed from these polygons can be monomially
equivalent, hence they lie in distinct monomial equivalence classes.
We do this by applying results from \S 1 to show that the minimum
distances (or in some cases, other parts of the complete weight enumerators)
are distinct.

\begin{thm} Let $q > 5$.  No two of the toric codes $C_P(\F_q)$ constructed
from the polygons in Theorem 3.5 are monomially equivalent.
\end{thm}

\begin{proof} If the dimensions are different, the toric codes
are certainly not monomially equivalent.  Hence we only need
to consider each $k$ separately.  

For the two codes with $k = 3$, Theorem 2.4 (which applies when $\ell = 0$ also)
shows $d(C_{P_3^{(1)}}) = (q-1)^2 - 2(q-1)$.  On the other hand,
Theorem 2.9 gives $d(C_{P_3^{(2)}}) = (q-1)^2 - (q-1)$.  Hence these
two codes are not equivalent.

For the codes with $k = 4$, Theorem 2.4 
shows $d(C_{P_4^{(1)}}) = (q-1)^2 - 3(q-1)$.  Corollary 2.10 shows
$d(C_{P_4^{(2)}}) = (q-1)^2 - 2(q-1)$.  Theorem 2.4 applies to
$C_{P_4^{(3)}}$ also, and shows $d(C_{P_4^{(3)}}) = ((q-1) - 1)^2 = (q-1)^2 - (2q - 3)$.
Finally, we must analyze $d(C_{P_4^{(4)}})$.  
Write $C_{P_4^{(4)}}(\F_q) = C(\F_q)$.  In this case, some more
advanced tools are needed.  If we translate this polygon 
by $(1,1)$ to place it in $\Box_{q-1}$, then we are evaluating
polynomials in $\mbox{Span}\{1,xy,x^2y,xy^2\}$ to get the codewords
of the corresponding (monomially equivalent) code.  Any linear combination 
of these monomials in which the coefficient of $x^2y$ or $xy^2$ is nonzero 
defines an absolutely irreducible curve of degree 3, whose 
closure in $\P^2$ has arithmetic genus $1$ by Theorem 4.2 of \cite{bp}.  
Hence by the general version of the Hasse-Weil bound from \cite{ap}, 
there can be at most $1 + q + 2\sqrt{q}$
$\F_q$-rational points on the corresponding affine curve.  
This means that the minimum distance of $C$ is 
at least $(q-1)^2 - (1 + q + 2\sqrt{q})$.  On the other hand, 
the other $k = 4$ examples have minimum distance no larger
than  $d(C_{P_4^{(3)}}) = (q-1)^2 - (2q - 3)$.  
It is easy to see from the quadratic formula that
$(1 + q + 2\sqrt{q}) < 2q - 3$ for all $q > 11$.  Hence
$d(C)$ is strictly larger than any of the others for
$q > 11$.  For the remaining small values of $q$ we check
directly that $d(C)$ is different from the others
using the Magma code described in \cite{j}.  The results 
are:
\begin{center}
\begin{tabular}{c|c|c}
 $q$ & $d(C(\F_q))$ & $(q-1)^2 - (2q-3)$ \\ \hline
      7 & 27 & 25\\ 
      8 & 40 & 36\\ 
      9 & 52 & 49\\ 
      11 & 85 & 81\\ 
\end{tabular} \\
$\quad$ \\
Table 1.  $d(C_{P_4^{(4)}}(\F_q))$
\end{center}
These are also different from any of the other $k = 4$ codes
over those fields.

For the $k=5$ codes, Theorem 2.4 
shows $d(C_{P_5^{(1)}}) = (q-1)^2 - 4(q-1)$.  Corollary 2.10 shows
$d(C_{P_5^{(2)}}) = (q-1)^2 - 3(q-1)$.  $C_{P_5^{(3)}}$ is a subcode
of the code $C_{P_2^{(2)}}$ from Theorem 2.9, and contains
codewords of the same minimum weight as the supercode. So
$d(C_{P_5^{(3)}}) = (q-1)^2 - 2(q-1)$.  The other three $k = 5$
codes also have $d(C_{P_5^{(i)}}) = (q-1)^2 - 2(q-1)$, which 
can be seen, for example, using Minkowski-decomposable subpolytopes
(the sets of three collinear points) as in \cite{ls}.  In 
$C_{P_5^{(4)}}$, for example, we have codewords 
$\mbox{ev}(b(xy - a_1)(xy - a_2))$, where $a_1,a_2,b \in \Fqstar$
and $a_1\ne a_2$, which have
$2(q-1)$ zeroes in $(\Fqstar)^2$.  

To show that the four codes with $d = (q-1)^2 - 2(q-1)$ are
not equivalent, we need to look at finer invariants.  
For instance, $C_{P_5^{(5)}}$ can be distinguished
from the other three by the {\it number of words} of minimum 
weight.  In $P_5^{(5)}$, there are 
two different sets of three collinear lattice points
while in the others, there is only one.  This means 
that there will be more words of the minimum 
weight in $C_{P_5^{(5)}}$ than in $C_{P_5^{(i)}}$
for $i = 3,4,6$.   
$C_{P_5^{(5)}}$ has at least $2{q-1\choose 2}(q-1)$ such words
because there are two distinct families of reducible 
polynomials: $b(x - a_1)(x - a_2)$ with $b,a_i \in \Fqstar$ and 
$a_1 \ne a_2$ and $b(y-a_1)(y^{-1}-a_2)$ $b,a_i \in \Fqstar$ and 
$a_1 \ne {a_2}^{-1}$.  
On the other hand, $C_{P_5^{(i)}}$
for $i = 3,4,6$ have (at least) ${q-1\choose 2}(q-1)$ such words.
There are more for some small $q$, but never as many as
$2{q-1\choose 2}(q-1)$.  See the weight enumerators
for $C_{P_5^{(6)}}$ over $\F_{11}$ and $\F_{16}$ in Table 2 below.
For sufficiently large $q$, we claim in fact that there are 
exactly ${q-1\choose 2}(q-1)$ such words.  This follows from 
the general Hasse-Weil bound from \cite{ap}.  For instance, for
$C_{P_5^{(6)}}$, if $q$ is
sufficiently large, then we claim all words in 
$C_{P_5^{(6)}}$ of weight $(q-1)^2 - 2(q-1)$
come from evaluations $\mbox{ev}(b(y-a_1)(y-a_2))$.
Any other such word could come only from evaluating a linear 
combination
of $\{1,x,y,y^2,x^{-1}y^{-1}\}$ in which $y^2,x,x^{-1}y^{-1}$
all appear with nonzero coefficients (since otherwise we are in a 
case previously covered).  Any such curve
is absolutely irreducible, of arithmetic genus 2
(because of the 2 interior lattice points in this case,
see Theorem 4.2 of \cite{bp}).  A simple argument shows that
$1 + q + 4\sqrt{q} < 2q - 2$ for all $q \ge 23$.
For smaller values of $q$, we verify directly
that the weight enumerators of $C_{P_5^{(5)}}(\F_q)$
do not match the weight enumerators of the other
codes using the Magma code from \cite{j}.  See Table 2.

To distinguish $C_{P_5^{(3)}}$ and $C_{P_5^{(4)}}$, we
use the codewords of weight $(q-1)^2 - (2q - 3)$ (one more than
the minimum weight).  Both of these codes contain
such words coming from evaluation of the polynomials
corresponding to the $1\times 1$ squares contained in these
polygons (copies of $P_4^{(3)}$).  Any such square
yields $(q-1)^3$ words of this weight since the polynomials
in question have the form $c(x-a)(y-b)$ and $a,b,c\in \Fqstar$
are arbitrary.  However $C_{P_5^{(4)}}$ has precisely
$(q-1)^3$ words of weight $(q-1)^2 - (2q - 3)$, while
$C_{P_5^{(3)}}$ has more of them, $3(q-1)^3$ to be specific.
This can be seen by considering the reducible
polynomials $d(x-a)(y-bx-c)$ that evaluate to give
codewords in $C_{P_5^{(3)}}$.  We get codewords
of weight $(q-1)^2 - (2q - 3)$ if $b = 0$, or if $c = 0$,
or if $b,c \ne 0$ and $a = -c/b$.

Finally, to distinguish $C_{P_5^{(6)}}$ from the other
three codes with $d = (q-1)^2 - 2(q-1)$, we must argue
as in the last case of the $k = 4$ codes.  If $q$ is
sufficiently large, then we claim $C_{P_5^{(6)}}$
contains no words at all of weight $(q-1)^2 - (2q - 3)$.
By Corollary 2.10 and the previous cases, we see that
any such word could come only from a linear combination
of $\{1,x,y,y^2,x^{-1}y^{-1}\}$ in which $y^2,x,x^{-1}y^{-1}$
all appear with nonzero coefficients.  As before, any such curve
is absolutely irreducible, of arithmetic genus 2. 
 A simple argument shows that
$1 + q + 4\sqrt{q} < 2q - 3$ for all $q > 23$.
Hence the Hasse-Weil bound from \cite{ap} shows
that there are no words of this weight for large $q$.  
For smaller values of $q$, we again verify directly
that the weight enumerators of $C_{P_5^{(6)}}(\F_q)$
do not match the weight enumerators of the other
codes.  See Table 2 below.

The following table gives the first three nonzero terms in
the weight enumerators:
$$W_C(x) = \sum_{i=0}^{(q-1)^2} A_i x^i,$$
where $A_i = |\{ w \in C : \mbox{wt}(w) = i\}$, 
for the $k = 5$ toric codes with $d = (q-1)^2 - 2(q-1)$.  
These were all computed using the Magma code from \cite{j}.

\bigskip
Over $\F_7$:
$$\begin{array}{cc}
      P_5^{(3)} & 1 + 90 x^{24} + 648 x^{25} + \cdots \\
      P_5^{(4)} & 1 + 90 x^{24} + 216 x^{25} + \cdots  \\
      P_5^{(5)} & 1 + 180 x^{24} + 324 x^{26} + \cdots \\
      P_5^{(6)} & 1 + 90 x^{24} + 432 x^{26} + \cdots 
\end{array}$$

Over $\F_8$:
$$\begin{array}{cc}
      P_5^{(3)} & 1 + 147 x^{35} + 1029 x^{36}+ \cdots \\
      P_5^{(4)} & 1 + 147 x^{35} + 343 x^{36}+ \cdots \\
      P_5^{(5)} & 1 + 294 x^{35} + 343 x^{37}+ \cdots \\
      P_5^{(6)} & 1 + 147 x^{35} + 1029 x^{37}+ \cdots 
\end{array}$$

Over $\F_9$:
$$\begin{array}{cc}
      P_5^{(3)} & 1 + 224 x^{48} + 1536 x^{49}+ \cdots \\
      P_5^{(4)} & 1 + 224 x^{48} + 512 x^{49}+ \cdots \\
      P_5^{(5)} & 1 + 448 x^{48} + 512 x^{51}+ \cdots \\
      P_5^{(6)} & 1 + 224 x^{48} + 512 x^{50}+ \cdots 
\end{array}$$

Over $\F_{11}$:
$$\begin{array}{cc}
      P_5^{(3)} & 1 + 450 x^{80} + 3000 x^{81}+ \cdots \\
      P_5^{(4)} & 1 + 450 x^{80} + 1000 x^{81}+ \cdots \\
      P_5^{(5)} & 1 + 900 x^{80} + 1500 x^{84}+ \cdots \\
      P_5^{(6)} & 1 + 650 x^{80} + 1000 x^{82}+ \cdots 
\end{array}$$

Over $\F_{13}$:
$$\begin{array}{cc}
      P_5^{(3)} & 1 + 792 x^{120} + 5184 x^{121}+ \cdots \\
      P_5^{(4)} & 1 + 792 x^{120} + 1728 x^{121}+ \cdots \\
      P_5^{(5)} & 1 + 1584 x^{120} + 7776 x^{126}+ \cdots \\
      P_5^{(6)} & 1 + 792 x^{120} + 1728 x^{125}+ \cdots
\end{array}$$

Over $\F_{16}$:
$$\begin{array}{cc}
      P_5^{(3)} & 1 + 1575 x^{195} + 10125 x^{196}+ \cdots \\
      P_5^{(4)} & 1 + 1575 x^{195} + 3375 x^{196}+ \cdots \\
      P_5^{(5)} & 1 + 3150 x^{195} + 13500 x^{203}+ \cdots \\
      P_5^{(6)} & 1 + 2250 x^{195} + 13500 x^{203} + \cdots \\
\end{array}$$

Over $\F_{17}$:
$$\begin{array}{cc}
      P_5^{(3)} & 1 + 1920 x^{224} + 12288 x^{225}+ \cdots \\
      P_5^{(4)} & 1 + 1920 x^{224} + 4096 x^{225}+ \cdots \\
      P_5^{(5)} & 1 + 3840 x^{224} + 5120 x^{232}+ \cdots \\
      P_5^{(6)} & 1 + 1920 x^{224} + 4096 x^{230}+ \cdots
\end{array}$$

Over $\F_{19}$:
$$\begin{array}{cc}
      P_5^{(3)} & 1 + 2754 x^{288} + 17496 x^{289}+ \cdots \\
      P_5^{(4)} & 1 + 2754 x^{288} + 5832 x^{289}+ \cdots \\
      P_5^{(5)} & 1 + 5508 x^{288} + 32076 x^{298}+ \cdots \\
      P_5^{(6)} & 1 + 2754 x^{288} + 5832 x^{294} + \cdots 
\end{array}$$

Over $\F_{23}$:
$$\begin{array}{cc}
      P_5^{(3)} & 1 + 5082 x^{440} + 31944 x^{441}+ \cdots \\
      P_5^{(4)} & 1 + 5082 x^{440} + 10648 x^{441}+ \cdots \\
      P_5^{(5)} & 1 + 10164 x^{440} + 154396 x^{454}+ \cdots \\
      P_5^{(6)} & 1 + 5082 x^{440} + 21296 x^{450}+ \cdots 
\end{array}$$

\bigskip
\center{Table 2.}

\bigskip
\noindent
Hence the enumerators never coincide for these four
codes, even in exceptional cases for small $q$.
\end{proof}

\bibliographystyle{amsalpha}

\end{document}